\documentclass[conference,usenatbib]{basi}
\usepackage[british]{babel}
\usepackage[varg]{txfonts}
\usepackage{rotating}
\usepackage{dcolumn}
\begin{document}
\title[Spectroscopic investigation of SDSS J100921.40+375233.9]{Spectroscopic investigation of SDSS J100921.40+375233.9 selected from SDSS and GALEX photometry}
\author[T. \c{S}ahin et~al.]%
       {Timur \c{S}ahin$^{1,2,3}$\thanks{email: \texttt{timursahin@akdeniz.edu.tr}},
      David L. Lambert$^{3}$ and Carlos Allende Prieto$^{4,3}$\\
       $^1$Department of Space Science and Technologies, Akdeniz University,
       Antalya, 07058, TURKEY\\
       $^2$TUBITAK National Observatory, Akdeniz University Campus, Antalya, 07058,
       TURKEY\\
       $^3$Department of Astronomy and The W. J. McDonald Observatory, University
       of Texas, Austin, TX 78712, USA\\
       $^4$Instituto de Astrofisica de Canarias, 38205, La Laguna, Tenerife,
       SPAIN}

\pubyear{2011}
\volume{00}
\pagerange{\pageref{firstpage}--\pageref{lastpage}}

\date{Received \today}

\maketitle
%------------------------------------------------------------------------------%
% abstract and keywords                                                        %
%------------------------------------------------------------------------------%
\label{firstpage}

\begin{abstract}

\noindent In this study, we aim to reveal the nature of the Sloan Digital Sky Survey (SDSS) star: SDSS J100921.40+375233.9, suspected to have an extremely low metallicity
We observed this star at high spectral resolution and performed an abundance analysis. 
We derived the spectroscopic parameters $T_{eff}$\,=5820$\pm$125\,K, $\log g$\,=\,3.9$\pm$0.2, and $\xi_t$\,=\,1.1$\pm$0.5\,km s$^{-1}$.
The star is consistent with belonging to the thick disk.

\end{abstract}


%------------------------------------------------------------------------------%
% main text of the paper, using \section, \subsection, \subsubsection          %
%------------------------------------------------------------------------------%
\section{Introduction}\label{s:intro}

Making use of the tables of cross-matched sources between Galaxy Evolution Explorer
({\sc GALEX}) and {\sc SDSS} available at the Multimission Archive at the Space
Telescope ({\sc MAST}), we have identified stars that exhibit colors
of extremely metal-poor stars.

Figure 1. is a combination of {\sc GALEX} and {\sc SDSS} photometry, i.e., a
stellar density map in the plane of the $NUV-g$
and $z-g$ colors for a sample of some 200,000 stars and reveals two accumulations of
sources: one centered at $z-g=-0.47$ and $NUV-g=3.7$ and associated with moderately
metal-poor F and G main sequence and subgiant stars. The second cluster at a similar
$z-g$ but at $NUV-g\simeq 2.7$ is populated by  white dwarfs. We have used the $g$, $z$ and
$NUV$ colors of BD\,$+17$\,4708 to set the zero point of the $z-g$ and
$NUV-g$ scales for model fluxes from Kurucz's 
model atmospheres (Kurucz 1993).
For the photometry of BD\,+17\,4708, the fluxes are taken from Bohlin \& Gilliland (2004). After this calibration, 
the calculated colours for a similar subgiant in the range of
effective temperature between 5900--6400 K with
metallicities typical of the thin disk ([Fe/H]=0), the thick
disk ([Fe/H]$=-0.7$), halo ([Fe/H]=$-1.5$), and ultra-low
metallicity ([Fe/H]$=-4.5$) population, are shown in Fig.~1. with 
solid lines. The same models for a dwarf star ($logg=5$)
correspond to the dashed lines in the figure.

\begin{figure}[ht]
\centerline{\includegraphics*[width=11cm]{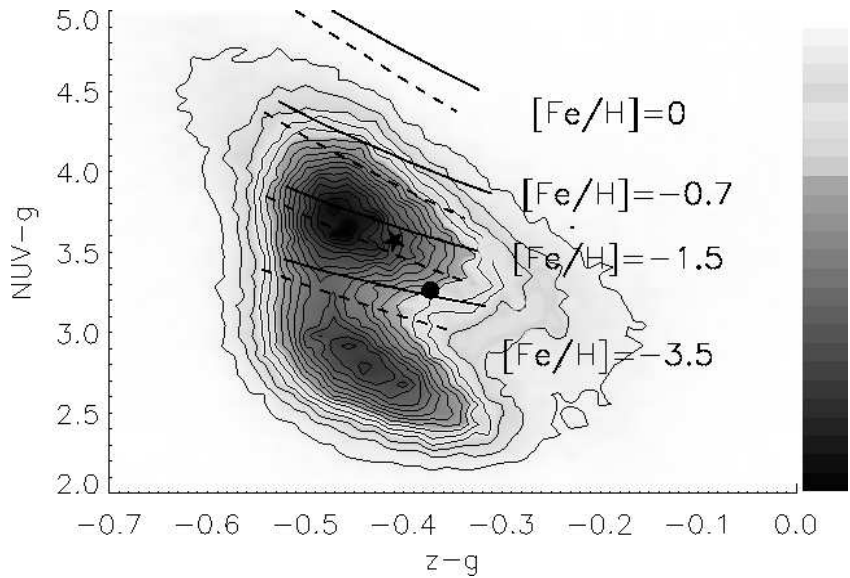}}
\caption{A stellar density map for a sample of 200,000
   stars which are photometrically similar to the flux standard BD+17\,4708
   (indicated with a star symbol). Filled circle represents J100921.}
\end{figure}

%------------------------------------------------------------------------------%
\section{Observations}\label{s:fonts}

\noindent High-resolution spectra for SDSS J100921 High-resolution spectra for SDSS J100921 (SDSS J100921.40+375233.9 = HIP 49750 = NLTT 23519) were obtained on 2009 March 15 (two frames) at the McDonald observatory with the 2.7 meter Harlan J. Smith reflector with the CCD-equipped Tull cross-dispersed \'{e}chelle spectrograph
(Tull et al. 1995). The spectra have a FWHM resolving power of 
$\lambda/\delta\lambda \simeq 60,000$ 
with full spectral coverage
from 3600 to 5300 \AA\ , and substantial but incomplete coverage from 5300 to
10\,200 \AA\ .

\noindent Observations were reduced using the echelle reduction package in {\sc
IRAF}. We refer the reader to \c{S}ahin et al. (2011) for details of data reduction.

\noindent We determined the heliocentric radial velocity of our target as V$_{\odot}$=$-$59.3$\pm$0.5 km s$^{-1}$. A section of the final spectrum is shown in Figure~2.

 \begin{table}
    \caption[]{Abundances of the observed species for SDSS J100921, J171422, and J015717 
are presented
    for the model atmospheres of $T_{\rm eff} = 5820$ K, $\log$ g = 3.9, $\xi$ = 1.1 and
    $T_{\rm eff} = 6320$ K, $\log$ g = 4.1, $\xi$ = 1.5, and $T_{\rm eff} = 6250$ K, $\log$ g = 3.7, $\xi$ = 4.0, respectively.
    The solar abundances from
    Asplund et al. (2009) are used to
    convert the abundance of element X to [X/Fe].}
       \label{}
   $$
       \begin{array}{l|ccc|c|c|ccc|c}
\hline
 $Species$     &  &J100921 &  & \log\epsilon_{\odot} &$Species$     &  &J100921 &  & \log\epsilon_{\odot}  \\
\cline{2-2}
\cline{3-3}
\cline{4-4}
\cline{5-5}
\cline{7-7}
\cline{8-8}
\cline{9-9}
\cline{10-10}
   &\log\epsilon(X) & $[X/H]$ & $[X/Fe]$ & & &\log\epsilon(X) & $[X/H]$ & $[X/Fe]$&   \\
 Li$\,{\sc i}$  &1.76 & 0.71&+2.01  &1.05&  Cr$\,{\sc i}$  &4.32 &-1.32&-0.02  &5.64 \\
 C$\,{\sc i}$   &7.58 &-0.85&+0.45  &8.43&  Cr$\,{\sc ii}$ &4.48 &-1.16&+0.14  &5.64 \\
 O$\,{\sc i}$   &8.28 &-0.41&+0.89  &8.69&  Mn$\,{\sc i}$  &3.91 &-1.52&-0.22  &5.43\\
 Na$\,{\sc i}$  &5.20 &-1.04&+0.26  &6.24&  Fe$\,{\sc i}$  &6.20 &-1.30&+0.00  &7.50\\
 Mg$\,{\sc i}$  &6.56 &-1.04&+0.26  &7.60&  Fe$\,{\sc ii}$ &6.20 &-1.30&+0.00  &7.50\\
 Al$\,{\sc i}$  &5.07 &-1.38&-0.08  &6.45&  Co$\,{\sc i}$  &3.98 &-1.01&+0.29  &4.99\\
 Si$\,{\sc i}$  &6.45 &-1.06&+0.24  &7.51&  Ni$\,{\sc i}$  &4.99 &-1.23&+0.07  &6.22\\
 Ca$\,{\sc i}$  &5.18 &-1.16&+0.14  &6.34&  Zn$\,{\sc i}$  &3.36 &-1.20&+0.10  &4.56\\
 Sc$\,{\sc ii}$ &2.06 &-1.09&+0.21  &3.15&  Sr$\,{\sc ii}$ &1.58 &-1.29&+0.01  &2.87\\
 Ti$\,{\sc i}$  &3.87 &-1.08&+0.22  &4.95&  Y$\,{\sc ii}$  &0.87 &-1.34&-0.04  &2.21\\
 Ti$\,{\sc ii}$ &4.08 &-0.87&+0.43  &4.95&  Zr$\,{\sc ii}$ &1.44 &-1.14&+0.16  &2.58\\
 V$\,{\sc i}$   &2.59 &-1.34&-0.04  &3.93&  Ba$\,{\sc ii}$ &0.98 &-1.20&+0.10  &2.18\\
 V$\,{\sc ii}$  &2.82 &-1.11&+0.19  &3.93&                 &     &     &       &    \\
\hline
       \end{array}
   $$
  \end{table}

\section{Spectral Analysis}

\noindent The abundance analysis was undertaken with models obtained by
interpolating in the ATLAS9 model atmosphere (ODFNEW models) grid (Castelli \& Kurucz 2003) and the line analysis programme MOOG (Sneden 2002). The models are line-blanketed plane-parallel atmospheres in Local
Thermodynamical Equilibrium (LTE) and hydrostatic equilibrium with flux (radiative plus convective)
conversation and computed with a constant microturbulent velocity
($\xi=$ 2 km s$^{-1}$).

\noindent The original sources for the transition probabilities of the Fe\,{\sc i} lines are listed by Lambert et al. (1996). The $gf$ values for Fe\,{\sc ii} lines are taken from Mel\'endez et al. (2006). 

\noindent The model parameters were determined using only spectroscopic criteria. Iron is is used to obtain an
estimate of log g. Finally, the metallicity [Fe/H] is refined by requiring that the derived abundance be equal to
that adopted for the construction of the model atmosphere for the final set of $T_{\rm
eff}$, $log\,g$, and $\xi$. For details of the spectral analysis, we
refer the reader to \c{S}ahin et al. (2012, in preperation). 

\noindent The model atmosphere parameters found for the star are $T_{eff}$\,=5820$\pm$125\,K, $\log g$\,=\,3.9$\pm$0.2, and $\xi_t$\,=\,1.1$\pm$0.5\,km s$^{-1}$

\begin{figure}
\centerline{\includegraphics[width=13cm]{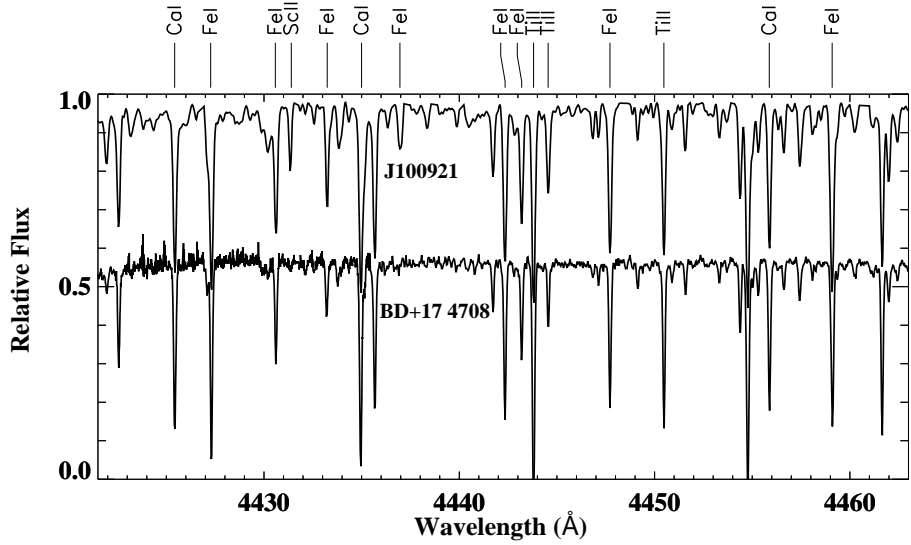}}
\caption{The velocity corrected spectra of SDSS J100921 and BD\,+17\,4708. Selected lines are identified.}
\end{figure}

\section{Concluding Remarks}

We performed the first detailed abundance analysis on SDSS J100921 for 21 elements. A summary of the abundances for the star is given in Table 1, where the quantities log$\epsilon$(X), [X/H], and [X/Fe] are reported in columns two, three, and four.

\noindent SDSS J100921.40+375233.9 appears to lie close to ultra-low metallicity
([Fe/H]=$-$3.5) population track (Fig~1.), with the fact that the SDSS
photometry for the star is saturated. Despite of the high risk, we have followed
spectroscopically the target to explore the nature of the star in detail. 

\noindent On the basis of preliminary results, we report that SDSS
J100921.40+375233.9 do not present ultra-low metallicity and is roughly
consistent with thick-disk membership. 

%------------------------------------------------------------------------------%
\section*{Acknowledgements}

I thank to Royal Astronomical Society (RAS) for providing the funds to attend the meeting.

%------------------------------------------------------------------------------%
% bibliography: produced from ADS using custom format of                       %
%                                                                              %
%     %z132 \\bibitem[%\2%(y)%\3m]%{R}\n   %\8.1g,%\Y,%\q,%\V,%\ p             %
%------------------------------------------------------------------------------%

\label{lastpage}
%------------------------------------------------------------------------------%

\begin{thebibliography}{}

\bibitem[\protect\citeauthoryear{Bohlin \& Gilliland}{2004}]{bohlin04} 
Bohlin R.~C., Gilliland R.~L., 2004, AJ, 128, 3053

\bibitem[\protect\citeauthoryear{Sneden}{2002}]{sneden02} 
Sneden, C., 2002, MOOG An LTE Stellar Line Analysis Program

\bibitem[\protect\citeauthoryear{Sahin et al.}{2011}]{sahin11}
\c{S}ahin T., Lambert D.~L., Klochkova V.~G., Tavolganskaya S., 2011,MNRAS, 410, 612

\bibitem[\protect\citeauthoryear{Tull et al.}{1995}]{tull95} 
Tull R.~G., MacQueen P.~J., Sneden C., Lambert D.~L., 1995, PASP, 107, 251

\end{thebibliography}
\end{document}